\documentstyle[amssymb,aps,pre,multicol,epsf]{revtex}

\begin{document}
\draft
\author{S.N. Dorogovtsev$^{1,2,*}$, J.F.F. Mendes$^{1,\dagger }$, and A.N. Samukhin$%
^{2,\ddag }$}
\title{Growing network with heritable connectivity of nodes}
\address{$^1$ Departamento de F\'\i sica and Centro de F\'\i sica do Porto, Faculdade 
de Ci\^{e}ncias, Universidade do Porto, \\ 
Rua do Campo Alegre 687, 4169-007 Porto, Portugal\\
$^2$ A.F. Ioffe Physico-Technical Institute, 194021 St. Petersburg, Russia}


\maketitle

\begin{abstract}
We propose a model of a growing network, in which preferential linking is
combined with partial inheritance of connectivity (number of incoming links) of
individual nodes by new ones. The nontrivial version of this model is solved
exactly in the limit of a large network size. We demonstrate, that the
connectivity distribution depends on the network size, $t$, in a
{\em multifractal} fashion. When the size of the network tends to infinity, 
the distribution behaves as $\sim q^{-\gamma }\ln q$, 
where $\gamma =\sqrt{2}$. For the finite-size network, this 
behavior is observed for $1 \ll q \lesssim \exp( \ln ^{1/2}t) $ 
but the multifractality is determined by the far wider part, $1 \ll q \lesssim \sqrt t$, 
of the distribution function. 
\end{abstract}


\begin{multicols}{2}

Hierarchically organized networks play an outstanding role in Nature. The well
known examples are the World Wide Web, scientific citations of papers,
networks of people's personal relations, neural networks, etc. (see, e.g.,
\cite{ws98,rs98,hppl98,ajb99,ba99,asbs00,nmej00,cmp00}). Most of these
networks belong to the class of {\em small world ones}, i.e., their diameters
are proportional to the logarithm of the network sizes \cite{ws98}. 
An important characteristic of a network is its connectivity distribution function, 
where the connectivity is the number of connections of a node (vertex degree, 
in the language of the graph theory). 
Properties of
networks depend dramatically on the form of this distribution. 
It has been realized recently, that growing networks with power-law
connectivity distributions, called {\em scale-free networks,} are of
special importance \cite{ba99,asbs00}. In particular, they are resilient to
random breakdowns \cite{ajb00,cnsw00,ceah00}.

A natural way to obtain a scale-free growing network is provided by the
mechanism of {\em preferential linking} \cite{ba99}. This principle is
similar to the one introduced in the well known Simon's model \cite{sha55}, used to explain 
power-law distributions in various social and economic systems. 
In fact, all these models belong to the class of stochastic multiplicative processes 
\cite{stoch}.
New links
are attached preferentially to nodes of a growing network with a high
number of connections. Several types of preferential linking were proposed,
which produced the $\gamma $ exponent of the connectivity distribution, $\Pi
\left( q\right) \sim q^{-\gamma }$ in the range $\left( 2,\infty \right)$
\cite{ba99,dms00,krl00}. Recently a model with $\gamma <2$ was also 
proposed \cite{dm00b}.

What are the scenarios of the evolution of networks produced by preferential
linking? What are the arising connectivity distributions? These questions
were considered in several recently published papers \cite
{dms00,krl00,dm00b,ab00}. However, in all the previous models, where the 
idea of preferential linking was used, it was
assumed, that new nodes appear with the same properties, independent of the
state of network at this moment, i.e., all nodes are born equal. One may say,
that new nodes are created by some invariable external source. Here we put forward
a different concept: {\em new nodes are born with random properties, which reflects
the state of the network at the moment of birth}. In this respect, one can
say, that they are created by the network itself. Of course, various
realizations of this idea are possible. For example, in \cite{sk00}, an
evolutionary model of such a type was proposed, but without preferencial 
linking. No power law distributions were found there.

Such inheritance of properties seems to be a rather usual feature of networks.
Let us discuss briefly, how it can arise in, e.g., networks of scientific
citations. In these networks, new nodes (scientific papers) arise not in
empty air. Each of them has its predecessors (e.g., some of the previous papers
of the same author or some papers on the same topic, etc.), and inherits a part
of their attractiveness. Roughly speaking, each paper is condemned to
popularity or oblivion already at the instant of its birth, and its future
depends on its direct predecessors. Similar inheritance may be essential in
collaboration networks, in networks of relations between firms in economy,
etc. Of course, a particular mechanism of the inheritance depends on a
particular network, but the idea of the inheritance is natural.

In the present Letter, we propose a model of a growing
network with {\em directed} links, in which preferential linking is combined
with partial inheritance of connectivity of individual nodes by new
ones. It is a simple example, demonstrating features of the network 
growth with such a combination of factors. The model is solved exactly in
the limit of large network sizes. This means that we do not pass to the continuous
limit of connectivity but use the {\em discrete} $q$ version of the model. 
Such an approach lets us to study features of the connectivity distribution 
which are invisible within continuous approximation.  

In the model, during the network growth, many
nodes appear and stay without any incoming connections. 
After exclusion of these nodes from the statistics, the average connectivity
(here it is a number of {\em incoming} links) of the remained part of the
network grows by a power law. The connectivity distribution becomes a
{\em multifractal} one. The body of the distribution 
($q \lesssim \exp(\ln^{1/2} q)$), which
includes most of the nodes, is of the form, $\Pi \left( q\right)
\sim q^{-\gamma }\ln t$, where $\gamma =\sqrt{2}<2$. The first
moment of connectivity is divergent, that means, that large fraction of links is concentrated ``in
hands'' of a small fraction of the nodes. 
Moreover, all the moments (except zero-th one) are  
determined by the behavior of the distribution   
in the region, $\exp[\ln^{1/2} t] \lesssim q \ll \sqrt t$.
Here the function is size-dependent but independent of the initial
conditions. Its form ensures that the moments scale with
the network size with exponents, nonlinearly dependent on their order. 
For $q \gtrsim \sqrt t$, 
the connectivity distribution is exponential.

{\em The model}.---At each increment of time, $dt$, a new node is
added with probability $dt$. At time $t$, the system consists of 
$N(t)$ nodes, 
$N\left( t\right)/t = 1 + {\cal O}(t^{-1/2})$. We
assume that a new node is born with a random number of incoming links, 
$q_N\left( t\right)$, which  
is distributed
according to some time-dependent distribution function, $\Pi _i\left( t,q_N\right)$. 
At the same time,
a link between sites is created with probability $mdt$. 
The probability that this link points at a node $j$ is $q_j\left( t\right) /Q\left( t\right) $, where $Q\left( t\right)
$ is the total number of links in the network, and $q_j\left( t\right) $ is
the connectivity of the node $j$, i.e., we use the same rule of preferential
linking, as in the Barab\'{a}si-Albert's model \cite{ba99}.
Since $Q(t)$ is an extensive variable, it may be replaced with $t\bar{q}\left( t\right) $, where $t\bar{q}\left( t\right)
=mt+\int^tdt^{\prime }\,\bar{q}_i\left( t^{\prime }\right) $. 
$\bar{q}_i(t)=\sum_q q \Pi_i(t,q)$.
The distribution
function of connectivity $\Pi \left( t,q\right) $ is defined as:

\begin{equation}
\Pi \left( t,q\right) =\left\langle \frac 1{N}%
\sum_{j=1}^{N}\delta \left[ q_j\left( t\right) -q\right]
\right\rangle \,\rightarrow \frac 1t\left\langle \sum_{j=1}^{N}\delta 
\left[ q_j\left( t\right) -q\right] \right\rangle .
\label{a1}
\end{equation}
Here, the average is over realizations of the process. 
Its infinitesimal variation, 
$d\Pi = d_n\Pi + d_l\Pi$, 
consists of two parts: 
$d_n\Pi =\left( dt/t\right) \left[ \Pi _i\left( t,q\right) -\Pi
\left( t,q\right) \right]$ 
arising from addition of new nodes and 
$d_l\Pi=\left[ mdt/t\bar{q}\left( t\right) \right] \left[ \left( q-1\right) \Pi
\left( t,q\right) -q\Pi \left( t,q\right) \right] $ 
originating from creation of new links.
Therefore, we get the following master equation:

\begin{eqnarray}
& & t\frac{\partial \Pi \left( t,q\right) }{\partial t}+\Pi \left( t,q\right) +%
\frac m{\bar{q}\left( t\right) }\left[ q\Pi \left( t,q\right) -\left(
q-1\right) \Pi \left( t,q-1\right) \right] 
\nonumber 
\\[5pt]
& & =\Pi _i\left( t,q\right) \,,
\label{1}
\end{eqnarray}

This equation is similar to the one used in \cite{dms00},
apart the term  $\delta \left( q-q_0\right) $ on the rhs replaced by $\Pi
_i\left( t,q\right) $.
Unlike the original Barab\'{a}si-Albert's model \cite{ba99}, a new node
is appeared not with predefined initial connectivity $q_i={\rm const}$, but $
q_i$ is assumed to be a random number, with a time-dependent probability
distribution function $\Pi _i\left( t,q\right) $.

We assume that $\Pi _i$ is some functional of the network properties, in
particular, on the connectivity distribution function, $\Pi $. In this respect,
{\em new sites are not created by some external source, but are born by the
network itself}. We propose the following inheritance rule for new nodes. We
assume, that every new node is born by some randomly chosen old one. At
the moment of birth it ``inherits''  
(copies), 
in average, a part $c$, $0<c<1$, of its
parent's connectivity. 
More precisely, with probability $c$, every of $k$ 
incoming links of a parent creates a link, pointing at its heir. The
parameter $c$ is, in turn, assumed to be a random number, distributed with
probability density $h\left( c\right) $. 
Hence, 

\begin{equation}
\Pi _i\left( t,q\right) =\int_0^1dc\,h\left( c\right)
\sum_{k=q}^\infty
{k \choose q}%
c^q\left( 1-c\right) ^{k-q}\Pi \left( t,k\right) 
\,.  
\label{2}
\end{equation}
From Eqs. (\ref{1}) and (\ref{2}) one can easily obtain:
$\bar{q}\left( t\right) =m/[(1-\bar{c})]\left( 1+bt^{\bar{c}-1}\right)$,
where $b$ is a constant of integration, and $\bar{c}$ is the average value of $
c$. Given we are interested mainly in networks of large sizes 
(equivalently, long times), in the following we set $b=0$. It is 
more convenient to rewrite Eqs. (\ref{1}) and (\ref{2}), using the Z-transformed
distribution function: $\Phi \left( t,y\right) =\sum_{q} \Pi \left( t,q\right) \left(
1-y\right) ^q$.
Then:

\begin{eqnarray}
& & t\frac{\partial \Phi \left( t,y\right) }{\partial t}+\Phi \left( t,y\right)
-\left( 1-\bar{c}\right) y\left( 1-y\right) \frac{\partial \Phi \left(
t,y\right) }{\partial y}-
\nonumber
\\[5pt]
& & \int_0^1dc\,h\left( c\right) \Phi \left(
t,cy\right) =0\,.  \label{5}
\end{eqnarray}
The initial condition is $\Phi \left(
t_0,y\right) =\Phi _0\left( y\right) $, $t_0\gg 1$ (this equation is
valid for $t\gg 1$). 
After rescaling of the size variable $t\rightarrow t/t_0$ the initial condition 
becomes $\Phi \left(1,y\right) = \Phi _0\left( y\right) $.

{\em Multifractality}.---The connectivity distribution in our model 
is of a {\em multifractal} type.
Indeed, the moments of the distribution may be expressed as: $%
M_n\left( t\right) =\sum_q\Pi \left( t,q\right) q^n=\left. [ \left(
y-1\right) \partial_y] ^n\Phi \left( t,y\right) \right| _{y=0}$. One
can easily derive from the equation (\ref{5}), that their dependence of
network size $t$ is 
$M_n\left( t\right) = A_{nn}t^{\tau \left( n\right) }+
A_{n,n-1}t^{\tau
\left( n-1\right) }+\cdots$ and $\tau (n) = \left( 1-\bar{c}\right) n+
\overline{c^n}-1\,,\;%
\overline{c^n}=\int\,dc\,c^nh\left( c\right)$.
The coefficients $A_{nk}$ depend on the initial distribution.
At long
times $M_n\left( t\right) \sim t^{\tau \left( n\right) }$, since 
$\tau \left( n\right) $ is a growing function for $n>1$. 
When $\tau \left( n\right) $ is not a linear function of $n$,
this type of 
size-dependent distribution is called multifractal \cite{mbb74,hp83,hjkps86}. 
(For a
pure fractal distribution it would be $M_n\left( t\right) \sim t^{\left(
n-1\right)D_f }$, where $D_f$ is the dimension of a fractal.) This function, $\tau \left(
n\right), $ encodes the set of generalized dimensions of the multifractal, $%
D_f\left( n\right) =\tau \left( n\right) /\left( n-1\right) $. 

Also, one may
describe a multifractal distribution as a statistical mixture of fractal ones with
``dimensionalities'', $\alpha =d\tau /dn$. These fractal distributions have
supports 
(sets of sites with non-zero connectivity)  
with dimensionalities, $f\left( \alpha \right) $, where 
$f\left( \alpha \right) $ is the Legendre transform of $\tau \left( n\right) $, $\tau
\left( n\right) +f\left( \alpha \right) =\alpha n$. In other words, every
fractal distribution enters with a statistical weight, scaling with network
size as $t^{-f\left( \alpha \right) }$. 

For example, if the distribution of $
c$ is uniform, $h\left( c\right) =\Theta \left( c\right) \Theta \left(
1-c\right) $, we obtain for our network:
$\tau(n) = n/2-n/(n+1)\,;\;D_f(n)=n/[2( n+1)]$, 
$\alpha =1/2-1/(n+1)^2$ 
and
$f(\alpha) = (1-\sqrt{1/2-\alpha})^2$ with 
$-\infty<\alpha <1/2$.

{\em The connectivity distribution}.---The distribution function 
$h(c)$ is specific for the given system, it
reflects its ``succession rights''. Here, we choose
this rule as: $h\left( c\right) =\Theta \left( c\right) \Theta \left(
1-c\right) $, i.e., $c$ is assumed to be uniformly distributed within the
interval $\left( 0,1\right) $. 
In this nontrivial and important case we are able to find the exact solution of the
problem. Indeed, for the homogeneous $h(c)$, after application of the
operator $\partial _y(y\cdot)$, Eq. (\ref{5}), 
may be reduced to a linear partial differential
equation. This equation, after the Mellin's transformation with respect to time, 
$\psi(\eta,y) = \int_1^\infty dt\, t^{\eta-1}\Phi(t,y)$, 
takes the form,

\begin{equation}
y^2\!\left( 1\!-\!y\right) \partial_y^2\psi + y\left( 2\eta -3y\right) 
\partial_y\psi
+ 2\eta \psi =\!-2\partial_y\!\left( y\Phi _0\right)\!
 ,  
\label{8}
\end{equation}
where $\Phi _0\left( y\right)$ is the initial distribution in the Z-representation.
In the following we shall consider the case $\Phi _0\left( y\right) =\left(
1-y\right) ^k$, that is $\Pi \left( 1,q\right) =\delta \left( q-k\right) $. 

The distribution, obtained with such initial condition, 
is denoted as $\Pi \left( t;q,k\right) $, and its Mellin's
time-transform --- as $\Psi \left( \eta ;q,k\right) $. 
Eq. (\ref{8}) may be
reduced to an inhomogeneous hypergeometric one after the substitution, $\psi \left( y\right)
=y^\zeta \chi \left( y\right) $, where $\zeta $ is one of the root of the
characteristic equation: 
$ \zeta ^2-\left( 1-2\eta \right) \zeta +2\eta =0 $.
Here we present the result in terms of 
$\Psi \left( \eta;q,k\right) $, which is the $q$-th term of Taylor's series of 
$\psi \left(y\right)$ 
around the point $y=1$. After lengthy calculations, we obtain 

\begin{eqnarray}
\Psi \left( \eta ;q,k\right) &=&\left\{
\begin{array}{l}
-\phi _1\left( \zeta _1,k\right) \frac{\phi _2\left( \zeta _1,q\right)
-\phi _2\left( \zeta _2,q\right) }{\zeta _1-\zeta _2}\,,\;k>q>0; 
\\[7pt]
-\frac{\phi _1\left( \zeta _1,k\right) -\phi _1\left( \zeta _2,k\right) }{%
\zeta _1-\zeta _2}\phi _2\left( \zeta _1,q\right) \,,\;q\ge \,k>0;
\end{array}
\right. 
\;  
\nonumber \\
\Psi \left( \eta ;0,k\right) &=&-\phi _1\left( \zeta _1,k\right) \;;\;\Psi
\left( \eta ;q,0\right) =-\frac{\delta \left( q\right) }\eta \;,  \label{10}
\end{eqnarray}
where $\zeta _1$ is the root of 
the characteristic equation,
which is positive for ${\rm Re}\,\eta <0$%
, $\zeta _2$ is the other root, and functions $\phi _{1,2}$ may be
expressed as: 
\end{multicols}
\widetext
\noindent\rule{20.5pc}{0.1mm}\rule{0.1mm}{1.5mm}\hfill

\begin{equation}
\phi _1\left( \zeta ,k\right) =
\frac{4k \Gamma (\zeta) \Gamma ( 2+\zeta )/(1-\zeta) }
{\Gamma \left( 1\!+\!\zeta +
\frac 2{1+\zeta }\right) \Gamma \left( 2\!+\!\zeta -\frac 2{1+\zeta }\right) }%
\!\int_0^1\!dz\,z^{1-2/\left( 1+\zeta \right) }\left( 1-z\right)
^{k-1}
F(\! 1-\frac 2{1\!+\!\zeta },1-\frac 2{1\!+\!\zeta };2+\zeta -\frac 2{%
1\!+\!\zeta };z) 
,  
\label{11}
\end{equation}
\vspace{-14pt}
\begin{equation}
\phi _2\left( \zeta ,q\right) = \frac{\Gamma \left( 1+\zeta +\frac 2{%
1+\zeta }\right) \Gamma \left( \frac 2{1+\zeta }-\zeta \right) }{\Gamma
\left( \frac 2{1+\zeta }-1\right) \Gamma \left( \frac 2{1+\zeta }+1\right)
}\frac{\sin \pi \zeta }\pi 
\int_0^\infty dy\,y^\zeta \left( 1+y\right) ^{-q-1}F\left( \zeta
,2+\zeta ;2+\zeta -\frac 2{1+\zeta };-y\right) 
\, .  
\label{12}
\end{equation}
Here $F$ is the hypergeometric function, $_2F_1$. 
\hfill\rule[-1.5mm]{0.1mm}{1.5mm}\rule{20.5pc}{0.1mm}
\begin{multicols}{2}
\narrowtext

Our main goal is to obtain $\Pi \left( t;q,k\right) $, 
at long $t$. One can calculate $\Pi$, the inverse Mellin's transform of $\Psi $, using the saddle point
approximation  
replacing the integration variable $\eta $ with $\zeta
\left( \eta \right) \equiv \zeta _1$.
At long $t$ the saddle point $\zeta _c$ is close to its
position at $t\rightarrow \infty $, $\zeta _\infty =\sqrt{2}-1$, 
$d\eta(\zeta\! =\!\zeta_\infty)/d\zeta = 0$. 
To avoid the situation when the integrand is zero in the saddle point, 
one may integrate the inverse Mellin's transform by parts. 
Then, in the saddle
point approximation at $t\rightarrow \infty $,   
we obtain the following relation,

\begin{equation}
\Pi \left( t;q,k\right) \approx 
\frac{t^{-3/2+\sqrt{2}}}{2^{1/4}\sqrt{\pi }%
\ln ^{3/2}t}\left. \frac{\partial \phi _1\left( \zeta ,k\right) }{\partial
\zeta }\frac{\partial \phi _2\left( \zeta ,q\right) }{\partial \zeta }%
\right| _{\zeta =\sqrt{2}-1} 
\, , 
\label{14}
\end{equation} 
for $q,k>0$. 

Note, however, that, if $t$ is finite, this expression becomes invalid for large enough $q$, 
since in this case, $\phi _2\sim q^{-\zeta }$ varies essentially within 
the saddle point maximum. 
 The region
of validity of Eq. (\ref{14}) may be estimated as $\ln q\ll \ln ^{1/2}t$. At
larger $q$, the saddle point position becomes $q$-dependent. For  $%
1\ll \ln q\lesssim \left( 1/2\right) \ln t$, the saddle point position remains
close enough to $\zeta _\infty $, and we obtain:

\begin{eqnarray}
& & \Pi \left( t;q,k\right) \approx 
\nonumber
\\[5pt]
& & 
d\frac{\ln \left(
aq\right) }{\left( t\ln t\right) ^{3/2}}\exp \left[ \sqrt{2\ln t\ln \left(
t/q^2\right) }\right] 
\left. \frac{\partial \phi _1\left( \zeta
,k\right) }{\partial \zeta }\right| _{\zeta =\sqrt{2}-1}
\,,  
\label{14a}
\end{eqnarray}
where: $d=0.174\dots$ and $a = 0.840 \dots$.

For non-zero initial connectivity $k$, the fraction of zero-connectivity nodes is

\begin{equation}
\Pi \left( t;0,k\right) \approx 1+\frac{t^{-3/2+\sqrt{2}}}{2^{1/4}\sqrt{\pi }%
\ln ^{3/2}t}\left. \frac{\partial \phi _1\left( \zeta ,k\right) }{\partial
\zeta }\right| _{\zeta =\sqrt{2}-1}
\,.  
\label{15}
\end{equation}

\begin{figure}
\epsfxsize=80mm
\epsffile{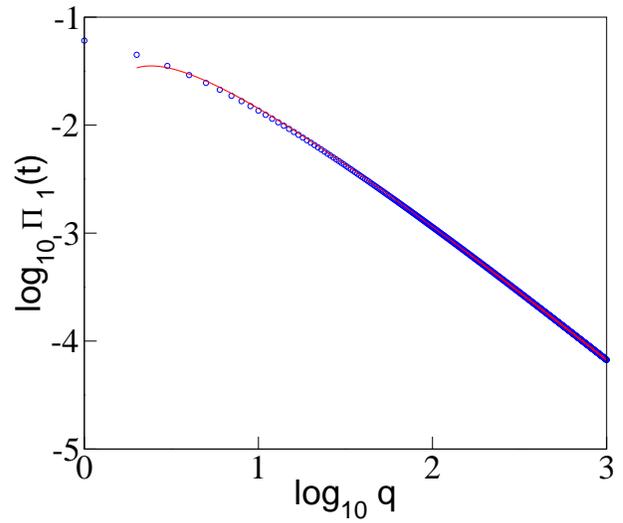}
\caption
{ Log-log plot of the stationary distribution, $\Pi _1\left( q\right)$, of incoming links.
The circles represent the exact result, Eq.~(\protect\ref{16}). 
The asymptotic dependence, Eq.~(\protect\ref{17}), is shown by the line. 
}
\label{fig1}
\end{figure}

One can see from Eq. (\ref{15}), that at long times, the fraction of
zero-connectivity nodes tends to $1$. These nodes do not have incoming links but  
only outgoing ones (remind, that connectivity, in the present paper, is defined as the number of incoming links). 
They are passive constituents of the network, i.e., their connectivity remains unchanged all the time.  
Although the
fraction of active nodes (with non-zero connectivity) tends to zero as the network grows, their total number, {\em increases} with time as $t^{\sqrt{2}-1/2}/\ln ^{3/2}t$. 
One can introduce the
distribution function of active nodes, $\Pi_1\left( t;q,k\right)$. 
It follows from Eqs. (\ref{14}) and (\ref{15}), that 
in the large time limit,
it tends to the distribution, $\Pi _1\left( q\right) $, which
depends neither on time nor on the initial conditions:

\begin{eqnarray}
\Pi _1\left( t,q\right) & \equiv & \left( 1-\delta _{q0}\right) \frac{\Pi \left(
t;q,k\right) }{1-\Pi \left( t;0,k\right) }\rightarrow 
\nonumber
\\[5pt]
& & \Pi _1\left( q\right) 
= -\left. \frac{\partial \phi _2\left( \zeta ,q\right) }{\partial \zeta }%
\right| _{\zeta =\sqrt{2}-1}
\,.  
\label{16}
\end{eqnarray}
A plot of $\Pi _1\left( q\right)$ is presented at Fig.\ref{fig1}.\ One can
obtain from Eqs. (\ref{16}) and (\ref{12}) the following asymptotic
expression for large $q$:

\begin{equation}
\Pi _1\left( q\right) =\frac d{q^{\sqrt{2}}}\ln \left( aq\right) 
\, .
\label{17}
\end{equation} 
Apart of the logarithmic factor, this is a power law distribution.
The exponent is less than $2$. Indeed, the first
moment of $\Pi _1\left( t,q\right) $ (average number of links per active
node) diverges at $t\rightarrow \infty$. 
At finite $t$, 
Eq. (\ref{17}) becomes invalid when $\ln q \gtrsim \ln ^{1/2}t$. One can
replace Eq. (\ref{17}) with more general expression, which follows from Eqs. (%
\ref{14a}) and (\ref{15}): 

\begin{equation}
\Pi _1\left( t,q\right) =dt^{-\sqrt{2}}\ln \left( aq\right) \exp \left[
\sqrt{2\ln t\ln \left( t/q^2\right) }\right] 
\, . 
\label{17a}
\end{equation} 
Here, $1 \ll q \ll k\sqrt{t}$.
Note, that the distribution in this region depends on the network size 
but not on the initial condition. 
Eq. (\ref{17}) follows from Eq. (\ref{17a}) in the appropriate region of $q$. 
Dependence on the initial connectivity distribution appears when $q \gtrsim k\sqrt{t}$.

One can also consider the limit, $q\rightarrow \infty $, keeping $t$ fixed. 
This determines the form of the large-connectivity cut-off of the scaling distribution. 
For $q-k\sqrt{t}\gg 1$, the
saddle point of the inverse Mellin's transform integral is shifted to the region 
$\eta <0$, $\left| \eta \right| \gg 1$. Evaluation of the integral yields 

\begin{equation}
\Pi \left( t;q,k\right) \approx \frac 1{\Gamma \left( k\right) t^{3/2}}%
\left( \frac {q-k}{\sqrt{t}}\right) ^k\exp \left( -\frac q{\sqrt{t}}\right) 
\label{18}
\end{equation} 
for $q \gtrsim  k\sqrt{t}$. 
The estimation of the total number of nodes for 
which the connectivity values are within this tail,  
$t_0 t \sum_{q=k\sqrt t}^{\infty}\Pi \left( t;q,k\right)$, 
gives a value of the order of $t_0$, 
the number of nodes in the initial state of the network.
The characteristic scale of the $q$ variation, $\sqrt t = t^{1-\bar{c}}$, 
can be obtained simply in the continuous $q$ limit of Eqs. (\ref{1}) and 
(\ref{2}) where the natural variable arises, $q/t^{1-\bar{c}} = q/\sqrt t$.   

{\em Conclusions}.---We have 
found the exact large-size solution of a growing network model, which
exhibits power-law type behavior, 
$\Pi(q) \sim q^{-\gamma}\ln q$, $1<\gamma<2$. 
In our model, $\gamma = \sqrt{2}$. 
With probability close to $1$, a randomly chosen node has its connectivity  
within the power law dependence region. In a recent paper
\cite{nmej00}, a number of real networks with $\gamma <2$ were presented. 
Note that the connectivity distributions with $\gamma<2$ were obtained 
analytically and by simulation for networks with accelerating growth \cite{dm00b}. 
In such networks, the average connectivity grows with the network size. 
In the present case, the number of links per {\em active} site 
also grows with time in a power law manner, i.e., 
as $t^{3/2-\sqrt{2}} \ln ^{3/2}t$.  
We have found that the characteristic feature of evolving networks with inheritance 
is the {\em multifractality} of the connectivity distribution.   
In our network, this intriguing property originates from the wide region, 
$\ln ^{1/2}t \lesssim q \lesssim  k\sqrt{t}$ of the connectivity distribution.

SND thanks PRAXIS XXI (Portugal) for a research grant PRAXIS
XXI/BCC/16418/98. JFFM was partially supported by the project
FCT-Sapiens 33141/99. We also thank A.V. Goltsev and B.N. Shalaev for many
useful discussions. 
{\em Note added}.---After this manuscript had been prepared, a paper of Bianconi and 
Barab\'{a}si \cite{bb00} has appeared, in which the multiscaling behavior 
and the connectivity distribution with logarithmic correction, 
$\Pi(q) \sim q^{-\gamma}/\ln q$, $\gamma>2$, was found for another model of growing networks. 
\\

\noindent
$^{*}$ Electronic address: sdorogov@fc.up.pt\newline
$^{\dagger }$ Electronic address: jfmendes@fc.up.pt\newline
$^{\ddag }$ Electronic address: alnis@samaln.ioffe.rssi.ru

\end{multicols}

\end{document}